\documentclass[aps,prb,preprint,showpacs,psfig,superscriptaddress,longbibliography]{revtex4-1}
\usepackage{epsfig}
\usepackage{graphics}
\usepackage{amsfonts}
\usepackage{mathrsfs}
\usepackage{amsmath}
\usepackage{color}
\usepackage{natbib}
\usepackage{textcomp}
\usepackage{graphicx}
\usepackage{braket}
\usepackage{bm}
\usepackage{amssymb}

\usepackage{xspace}
\usepackage{epstopdf}
\usepackage{dcolumn}
\usepackage{longtable}
\usepackage{subfigure}
\usepackage{multirow}
\usepackage[colorlinks=true, letterpaper=true, pdfstartview=FitV, linkcolor=blue, citecolor=blue, urlcolor=blue]{hyperref}

\makeatletter

\newcommand{\Rmnum}[1]{\expandafter\@slowromancap\romannumeral #1@}
\makeatother
\begin{document}
\title{Photonic Obstructed Atomic Insulator}
\author{Hongyu Chen}
\affiliation{Division of Physics and Applied Physics, School of Physical and Mathematical Sciences, Nanyang Technological University, Singapore 637371, Singapore}

\begin{abstract}
Topological quantum chemistry (TQC) classifies the topological phases by real-space invariants in which obstructed atomic insulators belong to the trivial case but sometimes show the feature of higher-order topological insulator. Here, for a two-dimensional magnetic photonic obstructed atomic insulator, we show that the emergence of corner states is associated with the Wyckoff positions. In such a square lattice with four +$M$ elements and four -$M$ elements, corner states appear when the superlattice exposes $2b$ Wyckoff positions. And the corner states will decrease when the number of exposed $2b$ Wyckoff positions decreases and disappears when $2b$ Wyckoff positions no longer stand at the edge. By arranging all the magnetic rods into one magnetization direction, we find that time-reversal symmetry is not important for corner states. Our finding indicates that the real-space distribution of atoms determines the feature of higher-order topological insulators for obstructed atomic insulators in TQC. 

\end{abstract}
\maketitle

Higher-order topological insulators (HOTIs) are the topological phases classified phenomenologically by bulk-boundary correspondence \cite{Xue2022}, where $d$-dimensional insulating bulk state is accompanied by ($d-k$)-dimensional ($k$$\textgreater$1) localized state. The quantized index of corner state is associate with the multipole moments of electric charge, where certain spatial symmetry will arrange fractional charge at the corners \cite{Benalcazar2017,Song2017}, namely, quadrupole state in two-dimensional (2D) system and octupole state in three-dimensional system as listed in Table. \ref{t1}:
\begin{table}[h]
\centering
\caption{\label{t1} Bulk-edge correspondence ($d$ - $k$, $k$ $\textgreater$ 1) in higher-order topological insulators.}
\begin{tabular}{c| c c}\hline\hline
$d$ & 2 & 3\\\hline
$k$ = 2 & corner state (quadrupole) & hinge state \\
$k$ = 3 & -& corner state (octupole) \\\hline\hline
\end{tabular}
\end{table}\\
These corner states defined by quantized multipole moments and hinge state were firstly realized in artificial system \cite{Imhof2018,Serra2018,Xue2019one,Xue2020,Peterson2018,He2020,Xie2020photon}. On the other hand, the discovery of HOTIs in condensed matter was realized by detecting the conducting step-edge states \cite{Schindler2018,Aggarwal2021}. Different from the first-order topological phase, HOTIs do not possess a unified index and the positions of localizes states rely more on the geometry such as crystal orientation \cite{Schindler2018}, exposed atomic configuration \cite{Hossain2024}.

Obstructed atomic insulators are the topological phases classified by the real-space topological invariant, topological quantum chemistry (TQC) \cite{Bradlyn2017,Robert2013,Jorrit2017}. In TQC, the elementary band representations (EBRs) are derived from the real-space site symmetry based on the symmetric Wannier functions ($s$, $p$, $d$, ... orbitals). Then for a lattice with certain symmetry with unknown topology, we can evaluate the first-order topology by the BRs at high-symmetry momentum points by the comparison with EBRs. In the first order topologically trivial case, the wavefunctions can be wannierized at the sites with and without the occupation of atoms, namely, atomic insulators (AIs) and obstructed atomic insulators (OAIs). The process of phase transition from AI to OAI is regarded as topologically nontrivial with the emergence of Dirac points and critical metal phase \cite{Wang2024}. The charge-filled atom-unoccupied Wyckoff positions (WPs) are called obstructed Wannier charge center (OWCC) \cite{Xu2024}. By cutting through OWCC in a 2D OAI, we will observe metallic edge state \cite{Wang2022,Hu2024} and sometimes corner states \cite{Wang2022}. Hence OAIs sometimes show the feature of HOTI which is dependent on the exposed atomic configuration. 


Previously, HOTIs have been applied in polariton lasing \cite{Wu2023}, nano laser \cite{Zhang2020}, multipolar lasing \cite{Kim2020} and plasmon launching \cite{Li2023}. And the flexibly-arranged resonators or waveguides in microwave band can enrich the theory for HOTIs such as bulk topology \cite{Serra2018}, double band inversion \cite{He2020}, sub-symmetry \cite{Wang2023}. The emergence of HOTI feature in OAI make it a good template for the investigation of confined light in photonic system.

We start from an eight-band tight-binding (TB) model:
\begin{equation}
\begin{aligned}
H_{8B} &= \tilde{t}\sum_{\braket{p,q}}c_{p}^{\dagger}c_{q} + t\sum_{\braket{\braket{p,q}}}e^{i\phi_{pq}}c_{p}^{\dagger}c_{q} + t^{\prime}\sum_{\braket{\braket{\braket{p,q}}}}c_{p}^{\dagger}c_{q}\\
\tilde{t} &= (\theta + \frac{\pi}{2})\frac{v}{a}; t = \frac{\sqrt{2} \pi v}{4a}; t^{\prime} = -\frac{\pi v}{4a}
\end{aligned}
\end{equation}
in which $p$, $q$ are the site indices, $\braket{.}$, $\braket{\braket{.}}$ and $\braket{\braket{\braket{.}}}$ represent the green (nearest neighbor), orange (square edges), and dashed black (square diagonals) bonds [see Fig. 1(a)]. And the red and blue dots in the diagram represent upward and downward magnetization. Particularly, this eight-band TB model is identified with a network model in Manhattan lattice, in which the scattering potential at the intersection ($C_2T$ center at WP $4c$) determines the transport behavior \cite{Wang2024}. The insulating bulk band is shown in Fig. 1(b). And we find double-degenerate in-gap corner states with similar real-space eigenfunctions [see Figs. 1(c) and 1(d)].

Photonic crystal (PhC) is the periodic arrangement of dielectric media \cite{Joannopoulos2008,Sakoda2004}. Particularly, time reversal symmetry could also be broken in PhC \cite{Haldane2008}. This is realized by the nonzero imaginary part in the permeability tensor of the gyromagnetic material:
\begin{equation}
\mu = \begin{pmatrix} 
\mu&i\kappa &0\\
-i\kappa& \mu&0\\
0& 0&\mu_0\\
\end{pmatrix}
\end{equation}
where $\mu_0$ is the vacuum permeability, $\mu$ and $\kappa$ are the parameters of the dielectric materials. 
To realize the theory in TQC via PhCs, COMSOL \cite{multiphysics1998introduction} is used. 

 I built a two-dimensional eight-element lattice via gyromagnetic material [see Fig. 2(a , e)]. The difference between the two unit cells in Fig. 2(a) and Fig. 2(e) are the distinguished exposed empty Wyckoff positions, $4c$ and $2b$. Both $4c$ and $2b$ are OWCC, by cutting which we can observe in-gap metallic edge states [see Figs. 2(c, d, g, h)]. However, we can only observe in-gap corner states in the former case [see Figs. 2(i-l)]. This cleaving-dependence emergence of HOTI is similar to the result in Bi and As \cite{Schindler2018,Hossain2024}. One possible reason is the symmetry of the whole square superlattice, $C_2T$ in the former case and $C_4$ in the latter case.

Referring to another square lattice with corner state \cite{Langbehn2017} where changing the crystal orientation will not violate the corner state even the reflection symmetry of the superlattice is broken, we also change the orientation from (100) [see Fig. 1(i)] to (110) [see Fig. 2(a) where the preserved symmetry is $C_4T$ and $C_2$ and the exposed Wyckoff positions are $4c$] and (120) [see Fig. 2(c) where the preserved symmetry is $C_2T$ and the exposed Wyckoff positions are $4c$ and $2b$]. It turns out that only the pattern with $C_2T$ symmetry possesses the in-gap corner state [see Figs. 2(b) and 2(d)] and the number of corner state is decreased. 

To further examine the necessity of $C_2T$ symmetry, we change the gyromagnetic ratio of the superlattice since changing magnetic field can trigger topological phase transition in photonic crystal. It turns out that the sign of gyromagnetic ratio will not impact the result of emergence of corner states [see Fig. 4]. This challenges the importance of $C_2T$ symmetry and indicates the importance of exposed atomic configuration.



\color{black}In this photonic OAI, we can see that cutting through an OWCC and preservation of $C_2T$ symmetry are the necessary and insufficient conditions for the emergence of corner states.\color{black} Cutting through WP $4c$ gives flat metallic double-degenerate edge states and double-degenerate corner states while cutting through WP $2b$ gives metallic edge states which are not flat and zero corner state. By rotating the square superlattice into (110) crystal orientation, we find that exposing WP $4c$ without considering crystal orientation cannot ensure the emergence of corner state. By rotating the square superlattice into (120) crystal orientation, we find that exposing WP $4c$ but keeping $C_2T$ symmetry can give one corner state. Finally, by changing the gyromagnetic ratio, we find that preservation of $C_2T$ symmetry is not the necessary condition for the emergence of corner state, indicating that crystal orientation and corresponding exposed atomic configuration are decisive. 






\bibliographystyle{unsrt}

\begin{thebibliography}{28}%
\bibitem{Xue2022} Haoran Xue, Yihao Yang, and Baile Zhang. Topological acoustics. Nature Reviews Materials, 7(12):974–990, 2022.
\bibitem{Benalcazar2017} Wladimir A. Benalcazar, B. Andrei Bernevig, and Taylor L. Hughes. Quantized electric multi-pole insulators. Science, 357(6346):61–66, 2017. 1 Haoran Xue, Yihao Yang, and Baile Zhang. Topological acoustics. Nature Reviews Materials,7(12):974–990, 2022.
\bibitem{Song2017} Zhida Song, Zhong Fang, and Chen Fang. (d-2)-dimensional edge states of rotation symmetry protected topological states. Physical Review Letters, 119(24):246402, 2017.
\bibitem{Imhof2018} Stefan Imhof, Christian Berger, Florian Bayer, Johannes Brehm, Laurens W. Molenkamp, Tobias Kiessling, Frank Schindler, Ching Hua Lee, Martin Greiter, Titus Neupert, and Ronny Thomale. Topolectrical-circuit realization of topological corner modes. Nature Physics, 14(9):925–929, 2018.
\bibitem{Serra2018} Marc Serra-Garcia, Valerio Peri, Roman S\"{u}sstrunk, Osama R. Bilal, Tom Larsen, Luis Guillermo Villanueva, and Sebastian D. Huber. Observation of a phononic quadrupole topological insulator. Nature, 555(7696):342–345, 2018.
\bibitem{Xue2019one} Haoran Xue, Yahui Yang, Guigeng Liu, Fei Gao, Yidong Chong, and Baile Zhang. Realization of an acoustic third-order topological insulator. Physical Review Letters, 122(24):244301, 2019.
\bibitem{Xue2020} Haoran Xue, Yong Ge, Hong-Xiang Sun, Qiang Wang, Ding Jia, Yi-Jun Guan, Shou-Qi Yuan, Yidong Chong, and Baile Zhang. Observation of an acoustic octupole topological insulator. Nature Communications, 11(1):2442, 2020.
\bibitem{Peterson2018} Christopher W. Peterson, Wladimir A. Benalcazar, Taylor L. Hughes, and Gaurav Bahl. A quantized microwave quadrupole insulator with topologically protected corner states. Nature, 555(7696):346–350, 2018.
\bibitem{He2020} Li He, Zachariah Addison, Eugene J. Mele, and Bo Zhen. Quadrupole topological photonic crystals. Nature Communications, 11(1):3119, 2020.
\bibitem{Xie2020photon} Biye Xie, Guangxu Su, Hong-Fei Wang, Feng Liu, Lumang Hu, Si-Yuan Yu, Peng Zhan, Ming-Hui Lu, Zhenlin Wang, and Yan-Feng Chen. Higher-order quantum spin hall effect in a photonic crystal. Nature Communications, 11(1):3768, 2020.
\bibitem{Schindler2018} Frank Schindler, Zhijun Wang, Maia G. Vergniory, Ashley M. Cook, Anil Murani, Shamashis Sengupta, Alik Yu Kasumov, Richard Deblock, Sangjun Jeon, Ilya Drozdov, H’el‘ene Bouchiat, Sophie Gu’eron, Ali Yazdani, B. Andrei Bernevig, and Titus Neupert. Higher-order topology in bismuth. Nature Physics, 14(9):918–924, 2018.
\bibitem{Aggarwal2021} Leena Aggarwal, Penghao Zhu, Taylor L. Hughes, and Vidya Madhavan. Evidence for higher-order topology in bi and bi0.92sb0.08. Nature Communications, 12(1):4420, 2021.
\bibitem{Hossain2024} Md Shafayat Hossain, Frank Schindler, Rajibul Islam, Zahir Muhammad, Yu-Xiao Jiang, Zi-Jia Cheng, Qi Zhang, Tao Hou, Hongyu Chen, Maksim Litskevich, Brian Casas, Jia-Xin Yin, Tyler A. Cochran, Mohammad Yahyavi, Xian P. Yang, Luis Balicas, Guoqing Chang, WeishengZhao, Titus Neupert, and M. Zahid Hasan. A hybrid topological quantum state in an elemental solid. Nature, 628(8008):527–533, 2024.
\bibitem{Bradlyn2017} Barry Bradlyn, L. Elcoro, Jennifer Cano, M. G. Vergniory, Zhijun Wang, C. Felser, M. I. Aroyo, and B. Andrei Bernevig. Topological quantum chemistry. Nature, 547(7663):298–305, 2017.
\bibitem{Robert2013}Robert-Jan Slager, Andrej Mesaros, Vladimir Juri\v{c}i\'{c} \& Jan Zaanen, The space group classification of topological band-insulators. Nature Physics, 9(2): 98-102, 2013.
\bibitem{Jorrit2017}Jorrit Kruthoff, Jan de Boer1, Jasper van Wezel1, Charles L. Kane, and Robert-Jan Slager. Topological Classification of Crystalline Insulators through Band Structure Combinatorics. Physical Review X, 7(4): 041069, 2017.
\bibitem{Wang2024} Fa-Jie Wang, Zhen-Yu Xiao, Raquel Queiroz, B. Andrei Bernevig, Ady Stern, and Zhi-Da Song. Anderson critical metal phase in trivial states protected by average magnetic crystalline symmetry. Nature Communications, 15(1):3069, 2024.
\bibitem{Xu2024} Yuanfeng Xu, Luis Elcoro, Zhi-Da Song, M. G. Vergniory, Claudia Felser, Stuart S. P. Parkin, Nicolas Regnault, Juan L. Ma˜nes, and B. Andrei Bernevig. Filling-enforced obstructed atomic insulators. Physical Review B, 109:165139, Apr 2024.
\bibitem{Wang2022} Lei Wang, Yi Jiang, Jiaxi Liu, Shuai Zhang, Jiangxu Li, Peitao Liu, Yan Sun, Hongming Weng, and Xing-Qiu Chen. Two-dimensional obstructed atomic insulators with fractional corner charge in the M A2Z4 family. Physical Review B, 106:155144, Oct 2022.
\bibitem{Hu2024} Jingnan Hu, Fei Yu, Aiyun Luo, Xiao-Hong Pan, Jinyu Zou, Xin Liu, and Gang Xu. Chiral topological superconductivity in superconductor-obstructed atomic insulator-ferromagnetic insulator heterostructures. Physical Review Letters, 132:036601, Jan 2024.
\bibitem{Wu2023} Jinqi Wu, Sanjib Ghosh, Yusong Gan, Ying Shi, Subhaskar Mandal, Handong Sun, Baile Zhang, Timothy C. H. Liew, Rui Su, and Qihua Xiong. Higher-order topological polariton corner state lasing. Science Advance, 9(21):eadg4322, 2023.
\bibitem{Zhang2020} Weixuan Zhang, Xin Xie, Huiming Hao, Jianchen Dang, Shan Xiao, Shushu Shi, Haiqiao Ni, Zhichuan Niu, Can Wang, Kuijuan Jin, Xiangdong Zhang, and Xiulai Xu. Low-threshold topological nanolasers based on the second-order corner state. Light: Science Applications, 9(1):109, 2020.
\bibitem{Kim2020} Ha-Reem Kim, Min-Soo Hwang, Daria Smirnova, Kwang-Yong Jeong, Yuri Kivshar, and Hong-Gyu Park. Multipolar lasing modes from topological corner states. Nature Communications, 11(1):5758, 2020.
\bibitem{Li2023} Yuanzhen Li, Su Xu, Zijian Zhang, Yumeng Yang, Xinrong Xie, Wenzheng Ye, Feng Liu, Haoran Xue, Liqiao Jing, Zuojia Wang, Qi-Dai Chen, Hong-Bo Sun, Erping Li, Hongsheng Chen, and Fei Gao. Polarization-orthogonal nondegenerate plasmonic higher-order topological states. Physical Review Letters, 130(21):213603, 2023.
\bibitem{Wang2023} Ziteng Wang, Xiangdong Wang, Zhichan Hu, Domenico Bongiovanni, Dario Juki$\prime$c, Liqin Tang, Daohong Song, Roberto Morandotti, Zhigang Chen, and Hrvoje Buljan. Sub-symmetry- protected topological states. Nature Physics, 19(7):992–998, 2023.
\bibitem{Joannopoulos2008} J.D. Joannopoulos. Photonic Crystals: Molding the Flow of Light (Second Edition). Princeton University Press, 2008.
\bibitem{Sakoda2004} K. Sakoda. Optical Properties of Photonic Crystals. Springer Berlin Heidelberg, 2004.
\bibitem{Haldane2008} F. D. M. Haldane and S. Raghu. Possible realization of directional optical waveguides in photonic crystals with broken time-reversal symmetry. Physical Review Letters, 100(1):013904, 2008.
\bibitem{multiphysics1998introduction} COMSOL Multiphysics. Introduction to comsol multiphysics®. COMSOL Multiphysics, Burlington, MA, accessed Feb, 9:2018, 1998.
\bibitem{Langbehn2017} Josias Langbehn, Yang Peng, Luka Trifunovic, Felix von Oppen, and Piet W. Brouwer. Reflection-symmetric second-order topological insulators and superconductors. Physical Review Letters, 119(24):246401, 2017.

\end{thebibliography}

\begin{figure*}[h]
\centering 
\includegraphics[scale = 1.1]{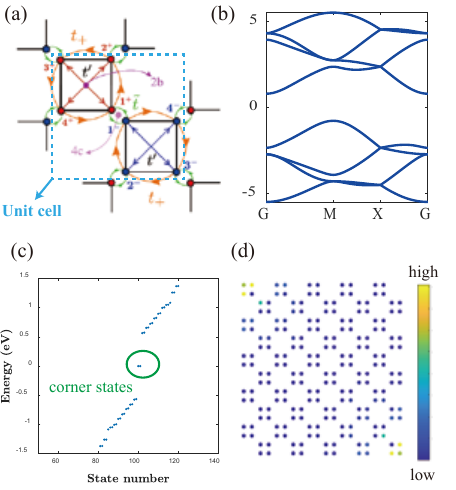}
\caption{Tight-binding model of a magnetic obstructed atomic insulator.  (a) The unit cell and hopping of the eight-band tight-binding model. (b) Band structures with $\theta$ = $\frac{\pi}{2}$. (g) Eigenvalues of 5$\times$5 supercell in which there are two corner states. (d) Distribution of eigenfunction of the corner states in real space. }    
\label{f1} 
\end{figure*}
\begin{figure*}[h]
\centering 
\includegraphics[scale =1]{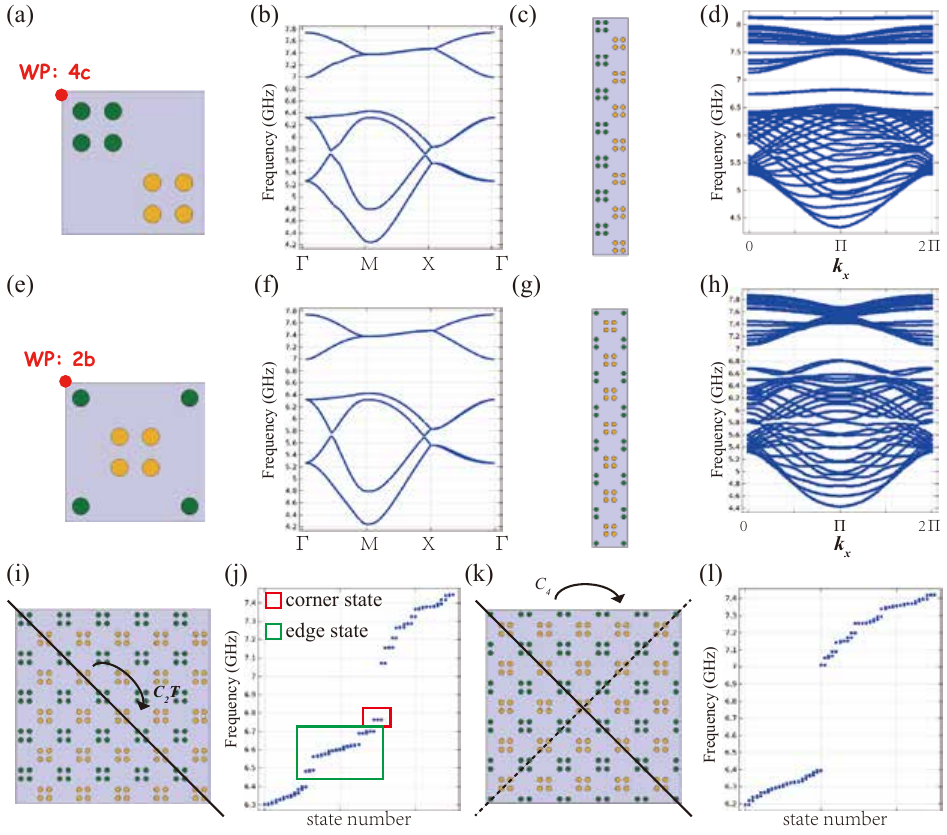}
\caption{Numerical simulation of photonic obstructed atomic insulator via COMSOL. (a, e) Unit cells and components of the PhC with the exposed sites of Wyckoff positions $4c$ and $2b$. The green and yellow sites represent YIG robs with positive and negative magnetization. (b, f) Band structures of the PhC (a, e). (c, g) Ribbons of PhC (a, e) where the boundary along $x$ direction is periodic and the boundary along $y$ direction is perfect electrical conductor. (d, h) Dispersion of edge states of (c, g). (i, k) Superlattice of PhC (a, e) with $C_2T$ and $C_4$ symmetry, respectively where the boundaries along $x$ and $y$ directions are both perfect electrical conductors.  (j, k) Simulated eigenvalues of the PhC (i, k) in which there are corner states in (j). } 
\label{f3} 
\end{figure*}

\begin{figure}[h]
\centering 
\includegraphics[scale = 0.8]{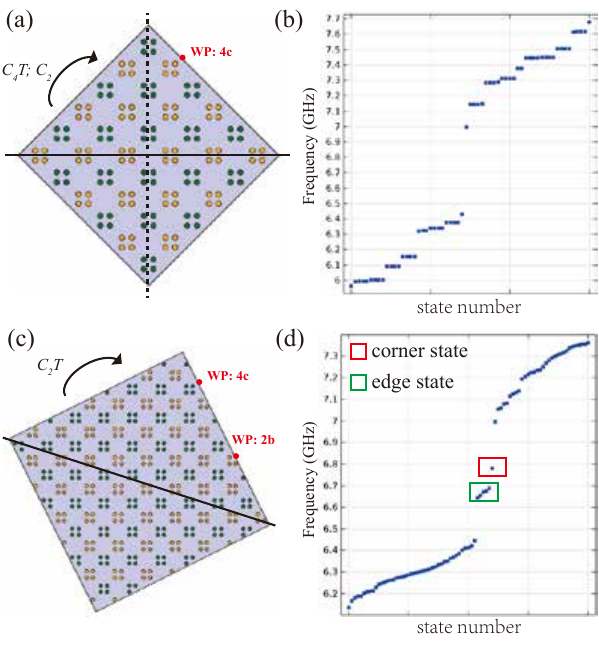}
\caption{Evaluating the robustness of the corner states by changing the crystal orientation. (a) Square superlattice along (110) crystal orientation. (b) Frequency spectrum of (a) with respect to state number. (c) Square superlattice along (120) crystal orientation. (d) Frequency spectrum of (c) with respect to state number. } 
\label{f3} 
\end{figure}
\begin{figure*}[h]
\centering 
\includegraphics[scale = 1]{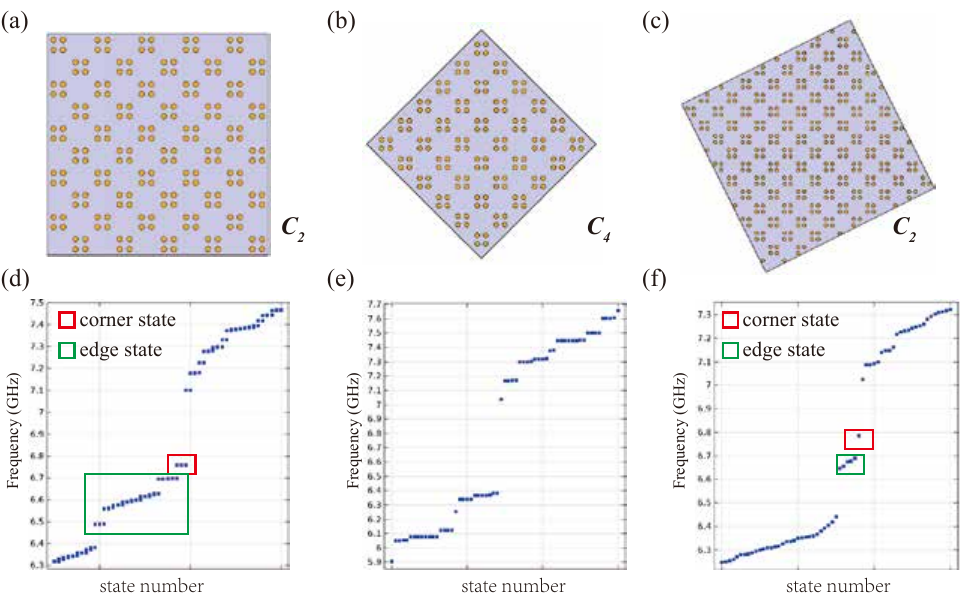}
\caption{Evaluating the robustness of the corner states by changing the gyromagnetic ratio and making all sites with positive magnetization. (a) Square superlattice with (a) (100) crystal orientation, (b) (110) crystal orientation and (c) (120) crystal orientation. (d), (e) and (f) Frequency spectra of (a), (b) and (c).} 
\label{f3} 
\end{figure*}

\end{document}